\documentclass[preprint,authoryear]{elsarticle}
\usepackage{graphicx}
\usepackage{color}

\newcommand{\fermi}{{\it Fermi} LAT}
\newcommand{\degr}{^\circ}
\newcommand{\fdg}{^\circ.}
\newcommand{\hess}{{H.E.S.S.}}

\begin{document}


\title{Origin of  Hydrogen Ionization in the 1 pc Galactic Central Region}

\author[tamm,hku]{D.O.~Chernyshov\corref{cor1}}
\ead{chernyshov@dgap.mipt.ru}
\author[hku]{Gene~C.K.~Leung}
\author[hku]{K.S.~Cheng}
\author[tamm,hku,mipt]{V.A.~Dogiel}
\author[par]{V.~Tatischeff}

\cortext[cor1]{Corresponding author}
\address[tamm]{I.E.Tamm Theoretical Physics Division of P.N.Lebedev
Institute of Physics, Leninskii pr. 53, 119991 Moscow, Russia}
\address[hku]{Department of Physics,
University of Hong Kong, Pokfulam Road, Hong Kong, China}
\address[mipt]{Moscow Institute of Physics and Technology, 141700 Moscow Region, Dolgoprudnii, Russia}
\address[par]{Centre de Sciences Nucl\'eaires et de Sciences de la Mati\'ere,
IN2P3/CNRS and Univ Paris-Sud, 91405 Orsay Campus, France}

\begin{abstract}
We study a possible connection between processes of gamma-ray
emission  and hydrogen ionization in  a few pc of central region around Sgr A*.  Previous
investigations showed there is a discrepancy between
interpretation of gamma-ray and ionization data if gamma-rays are generated by proton-proton collisions.  Here we provided analysis of processes of ionization and emission basing on analytical and numerical calculations of kinetic equations which describe processes of particle propagation and their energy losses. We assumed that cosmic rays (CRs) are emitted by a central source. The
origin of gamma rays could be either due to collisions of
relativistic protons with the dense gas of the surrounding
circumnuclear disk (CND) or bremsstrahlung and inverse
Compton scattering of relativistic electrons. The hydrogen ionization in this case is produced  by a
low energy component of the CR spectrum.   We found that if ionization is produced by protons  the expected ionization rate of hydrogen in the CND is of the same order  as derived from IR observations. So we do not see any
discrepancy between the gamma-ray and ionization data for the hadronic model. In the case of ionization by electrons we obtained the ionization rate one order of magnitude higher than follows from the IR data. In
principle,  a selection between the leptonic and hadronic
interpretations  can be performed basing on
measurements of radio and X-ray fluxes from this region because the leptonic
and hadronic models give different values of the fluxes from there.
 We do not exclude that  gamma-ray production
and hydrogen ionization in the CND are due to a past activity
of Sgr A* which occurred about 100 year ago. Then we  hypothesize that there may be
connection between a past proton eruption and a flux of hard X-rays emitted by Sgr A* hundred years ago as follows from the observed
 time variability of the iron line seen in the direction of GC molecular clouds.
\end{abstract}


\date{\today}

\maketitle

\section{Introduction}

The central region around Srg A* of the radius
about several pc is characterized by very peculiar parameters of
the interstellar medium \citep[see e.g.][]{katia07,katia12}. The
central black hole is surrounded by the circumnuclear disk (CND)
whose total mass was estimated by \citet{chr05} as $10^6M_\odot$.
The analysis performed by \citet{katia12} gives a slightly lower
value of $\sim 2\times 10^5M_\odot$ for the region of radius
$R_c=3-5$ pc that gives the average gas density in the CND of
about $n_H\simeq 4\times 10^5$ cm$^{-3}$.

\begin{figure}[h]
\begin{center}
\includegraphics[width=0.8\textwidth]{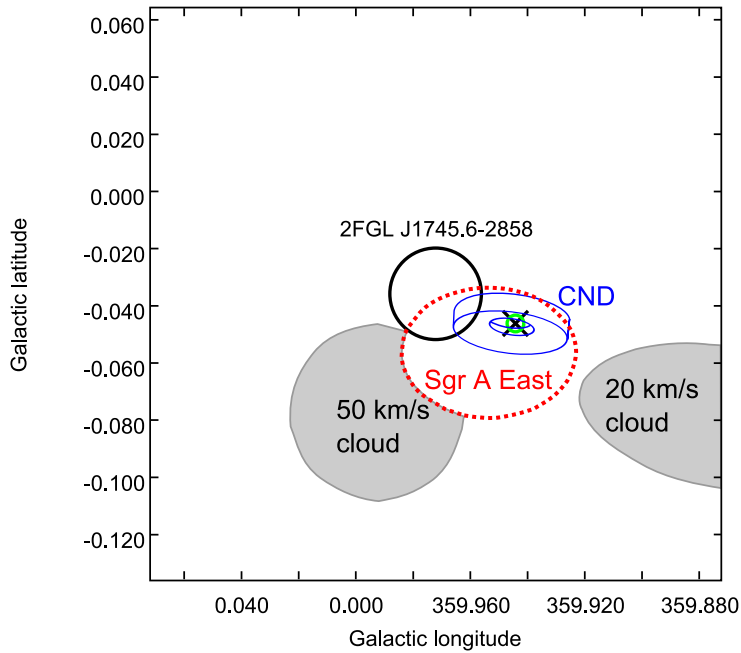}
\end{center}
\caption{Positions of sources within the
$0\fdg2\times0\fdg2$ region centered at 2FGL J1745.6$-$2858. The
$95\%$ confidence positional error circle of 2FGL J1745.6$-$2858 
in 2FGL is
labeled with a black circle, the position of Sgr A* with a black
cross and the position of the IRS stars with a green circle. The
 molecular complexes CND, 20 km/s and 50 km/s
\citep[see][]{katia12} are also shown in the figure.}
\label{fig:bgmap}
\end{figure}

Recent observations of the The High Energy Stereoscopic System
(\hess) \citep{aha09} and {\it Fermi}  Large Area Telescope
(\fermi) gamma-ray telescopes \citep[see][]{masha11}  found a
prominent gamma-ray flux in the TeV and GeV regions (the source
2FGL J1745.6$-$2858 in the second \fermi~ source catalog  \citep[see][]{nolan12})
that indicated that this region is filled with high
energy cosmic rays (CRs). The estimated gamma-ray flux  from  2FGL
J1745.6$-$2858 for $E>2$ GeV is about $I_{obs}=1.08\times 10^{-10}$ erg
cm$^{-2}$s$^{-1}$  with the spectral index $\gamma=2.68$ \citep[see][]{masha11} that corresponds a  luminosity  about
$8\times 10^{35}$ erg s$^{-1}$.  In Fig. \ref{fig:bgmap} the CND region is
shown
 by the red circle. The position of Sgr A* is shown  by the black
cross.  Positions of IR sources used by \citet{goto13} are shown in  Fig.
 \ref{fig:bgmap} by the green circle. The
$95\%$ confidence positional error circle of 2FGL J1745.6$-$2858 is
labeled with a black circle. The
 molecular complexes CND, 20 km/s, 50 km/s and the SNR Sgr A East
 are also shown in the figure. We derived the $95\%$ confidence positional error circle for  \fermi~ data
collected in the time interval  from August 4, 2008 to May 30,
2014. The data were reduced and analyzed using the {\it Fermi}
Science Tools package (v9r32p5), available from the {\it Fermi}
Science Support Center
\footnote{http://fermi.gsfc.nasa.gov/ssc/data/analysis/software/}.

Additional evidence for CRs in the GC was obtained by \citet{goto13,goto14}
 who  estimated the rate of hydrogen ionization
in the GC from observations of IR absorption lines in spectra of
several sources. Positions of IR sources used by \citet{goto13} are shown in  Fig. \ref{fig:bgmap} by the green circle. 
\citet{goto13} estimated the ionization rate in the 1 pc
region of the CND   as $\zeta \simeq 1.2\times 10^{-15}$ s$^{-1}$ which is of
the order of the ionization rate derived for the more extended
central region of radius $\sim$ 100 pc \citep[see,
e.g.,][]{oka05}. It was concluded that the ionization in the 100
pc region is mainly produced by subrelativistic CRs whose source
luminosity in the GC is no more than $10^{38}-10^{39}$ erg
s$^{-1}$ \citep[see, e.g.,][]{dog13,dog14,yus3}.

As seen from observations, the positional error circle of 2FGL
J1745.6$-$2858 overlaps with the CND, which is compatible with the
picture that its emission originates within the CND. It is
reasonable to assume that this emission is provided by CRs emitted
by  Sgr A* as assumed by \citet{masha11}.  The energy release in
Sgr A* may be large enough to produce this gamma-ray flux. Thus,
\citet{yusef12} concluded from high resolution X-ray and radio
data within 5 pc central region that Sgr A* released energy during
the last 1-3 hundred years in the form of a jet with the gamma
factor $\gamma\sim 3$. The estimated outflow kinetic luminosity is
about $1.2\times 10^{41}$ erg s$^{-1}$.

  \citet{goto13} analyzed whether the hydrogen ionization
can be provided by  CR protons which also generate there the
observed gamma-ray flux. The proton spectrum at $E>1$ GeV at the Galactic center
 was derived by \citet{masha11} in the assumption that the
gamma-ray flux from the source 2FGL J1745.6$-$2858 was produced by
proton-proton ($p-p$) collisions in the 3 pc CND region where the
gas density  equaled $\bar{n}_H=10^3$ cm$^{-3}$. Our
estimations show, that if the spectrum  is as derived by
\citet{masha11}, the ionization rate produced by relativistic
protons only  with  $E>1$ GeV is about $10^{-14}$ s$^{-1}$, 
i.e. one order of magnitude higher than the values of
\citet{goto13} obtained from the IR observations.

\citet{goto13}
assumed that if this spectrum can be extrapolated  into the range
$E<1$ GeV, the expected ionization rate, produced
by subrelativistic protons, was four orders of magnitude higher than
that deduced from the H$_3^+$ observations.

  All of these lead to a strong discrepancy
between the observed ionization rate within the CND region
and the hadronic  model of gamma-rays in the CND region.

Below we re-analyze the origin of GeV gamma rays and the hydrogen
ionization in the CND  in order to find an alternative explanation
of this discrepancy.

We re-analyze  processes of ionization and gamma-ray production in
the CND for the  hadronic and leptonic models, namely:
\begin{itemize}
\item We assume that CRs are emitted in the CND by a central source which can be either Sgr A* or any SNR e.g. Sgr A East;
\item Ionization and gamma rays are produced by CR protons. In this case gamma-ray photons are generated by $p-p$ collisions;
\item Ionization and gamma rays are produced by CR electrons. In this case gamma-ray photons are produced by
bremsstrahlung and inverse Compton scattering. The inverse Compton model of gamma-ray emission from the Sgr A* was presented e.g. in \citet{kusunose}. 
\end{itemize}

Unlike previous investigations we derive the spectrum of emitting particles in the relativistic and subrelativistic energy ranges from corresponding kinetic equations which include processes of energy losses and, if necessary, particle propagation. The parameters of CND region were taken from  \citet[see][]{katia12}.
The strength of magnetic field around Sgr A* and in the CND is about $(1-4)\times 10^{-3}$ G \citep[see][]{kill92, yusef96, eat13},
and the density of background photons (IR and optical) $w_{ph}\sim 10^4$ eV cm$^{-3}$ \citep[see][]{metz96}.

\section{\fermi~observations}
Previous \fermi~observations \citep[e.g.][]{masha11,nolan12} 
on this source made use of 2 years of data. Now with 6 years of 
\fermi~data available, it is useful to perform an up-to-date 
analysis to benefit from the increased statistics.
For $\gamma$-ray observation, we used {\it Fermi} LAT data
collected in the time interval  from August 4, 2008 to May 30,
2014. The data was reduced and analyzed using the {\it Fermi}
Science Tools package (v9r32p5), available from the {\it Fermi}
Science Support Center
\footnote{http://fermi.gsfc.nasa.gov/ssc/data/analysis/software/}.
We selected events in the reprocessed Pass 7 'Source' class and
used the P7REP\_SOURCE\_V15 version of the instrument response
functions. To reduce contamination from the gamma-rays produced in the upper atmosphere, we
excluded time intervals when the region of interest (ROI) was observed at zenith angles
greater than $100\degr$ or when the rocking angle of the LAT was
greater than $52\degr$. We used photons between 0.1 and 300 GeV
within a $20\degr\times20\degr$ ROI centered at the position of
2FGL J1745.6-2858.

We performed a binned likelihood analysis with the gtlike tool.
For source modeling,  all 2FGL catalog sources \citep[][]{nolan12}
within $19\degr$ of the ROI center, the galactic diffuse emission
(gll\_iem\_v05.fits) and isotropic diffuse emission
(iso\_source\_v05.txt) were included. For sources more than
$10\degr$ away from the ROI center, all spectral parameters were
fixed to the catalog values. We modeled 2FGL J1745.6-2858 with a
log-parabola function as in the 2FGL catalog
\begin{equation}
\frac{dN}{dE} = N_0 \left(\frac{E}{E_0}\right)^{-(\alpha+\beta\log(E/E_0))}.
\end{equation}

Spectral points were obtained by fitting the normalization factors
of point sources within $3\degr$ from the  source of interest and
the diffuse backgrounds in individual energy bins. Other
parameters were fixed to the best-fit values found in the full
energy band fit. In each energy bin, an initial fit was performed
using the DRMNFB optimizer, which has a weaker convergence
criterion. We identified and removed all point sources with TS
values $< 0$ in that energy bin, and re-optimized the fit using
the NEWMINUIT optimizer, which has a stricter convergence
criterion.

We note that the Galactic Center
region has a strong diffuse emission background and a high density
of point sources. A total of 10 2FGL sources are located within
$2\degr$ from the galactic center, and the nearest source 2FGL
J1746.6-2851c and 2FGL J1747.3-2858c are located only $0\fdg25$ 
and $0\fdg66$ from the source of
interest. This leads to a possibility of background confusion, which 
affects the lowest energy band most severely due to the large
point spread function of the LAT at $\sim100$ MeV.

We investigated the effect of the background confusion in the
100-158 MeV energy bin by performing two 
separate fits where the the fluxes of 2FGL
J1746.6-2851c and 2FGL J1747.3-2858c were fixed to the lower and
upper bounds obtained in the full energy fit, respectively. 
The resulting fluxes were $\sim20\%$ lower than that obtained in a
simple fit. We adopted these
two values as the lower and upper bounds of the flux in 100-158 MeV.

Figure \ref{fig:bgmap} shows the positions of the sources in a
$0\fdg2\times0\fdg2$ region centered at the position of 2FGL 
J1745.6-2858. The $95\%$ confidence positional
error circle of 2FGL J1745.6-2858 from the catalog and the size of the
circumnuclear disk are shown in the Figure (see figure captions).
The overlap between the source and the circumnuclear disk shows
that the emission may originate within the CND.

The spectrum of the gamma-ray emission is presented in Section 5. For comparison we also
plotted in this figure the data from \citet{masha11} and from 2FGL.
Our spectrum agrees with 2FGL in general, but the
increased statistics allowed for smaller error bars and finer binning in
energy, which could put better constraint for various models.

\section{Kinetic equation for CRs in the CND region}

 The general kinetic equation for CRs can be presented in the form
\begin{equation}
\frac{\partial N}{\partial t}-\nabla(D\nabla N)+\frac{\partial}{\partial E}\left(\frac{dE}{dt}N\right)+\frac{N}{T}=Q(E,r,t)
\label{eq_state}
\end{equation}
where $N(E)$ is the CR density, $E$ is the particle energy, $dE/dt$ is the rate of energy
losses for protons or electrons,  $Q(E)$ is the source function,
$T$ is the particle lifetime, which e.g. for protons is the time
of $p-p$ collisions, and $D$ is an affective diffusion coefficient
due to scattering on magnetic  fluctuations \citep[see
e.g.][]{ber90}.

If a point-like source emits a power-law momentum spectrum of
particles ($\propto p^{-\gamma}$) with the spectral index $\gamma$ then its energy spectrum can be presented as
\begin{equation}
Q(r,E,t)=A(E,t)\delta({\bf r})
\label{Q}
\end{equation}
with
\begin{equation}
\label{s_sp}
 A(E)=A_0(t){{E+Mc^2}\over{(E^2+2Mc^2E)^{(\gamma+1)/2}}}
\label{q_pwlaw}
\end{equation}
Here $M$ is the particle mass, $E$ its energy, $\gamma$ is the injection
spectral index of protons and $A_0$ is proportional to the
source power.

In the non-relativistic energy range the injection spectrum is
transformed by ionization losses. The rate of ionization losses is
\citep{haya, ginz}
\begin{equation}
\left({{dE}\over{dt}}\right)_i=-{{2\pi
e^4n_H}\over{m_ec\beta(E)}}\ln\left({{m_e^2c^2W_{max}} \over{4\pi
e^2\hbar^{2} n}}\right) \label{iot}
\end{equation}
where $n_H$ is the density of background gas, $m_e$ is electron mass, $W_{max}$ is the highest energy
transmitted to an ambient electron,  and $\beta(E)=v/c$.
The accurate expression for
energy losses at low energies was taken from  the PSTAR and ESTAR
database  \citep[see][]{star} for protons and electrons, for electron energies below 10 keV we used the approach of \citet{dalg}.

The process of catastrophic $p-p$ collisions can be presented as 
continuum energy losses, and  the approximate formula for $p-p$
energy losses can be given as \citep[see][]{mns94},
\begin{equation}
\left(\frac{dE}{dt}\right)_{pp}=-0.65cn_H\sigma_{pp}\theta\left(E-1.22~\mbox{GeV}\right)\,,
\end{equation}
where  $\sigma_{pp}$ is the cross-section of proton-proton collisions.
Thus for protons
\begin{equation}
\frac{dE}{dt} = \left(\frac{dE}{dt}\right)_{pp} + \left({{dE}\over{dt}}\right)_i \,.
\label{sloss}
\end{equation}

As one can see from Eq. (\ref{eq_state}) the problem is
characterized by the three times which for the CND  parameters
are:
\begin{itemize}
\item The characteristic time of $p-p$ collisions
\begin{equation}
\tau_{pp}=n_Hc\sigma_{pp}\simeq 100~\mbox{yr}\,;
\end{equation}
\item The characteristic time of ionization losses
\begin{equation}
\tau_i=\int\limits_E\frac{dE}{(dE/dt)_i}~\mbox{yr}\,,
\end{equation}
that gives $\tau_i\simeq 10^3\left(\frac{1~{GeV}}{E}\right)^{3/2}$ yr for subrelativistic protons;
\item The characteristic time of diffusion
\begin{equation}
\tau_D\simeq\frac{R^2}{4D}\,,
\end{equation}
where $R\simeq 3$ pc is the size of the CND region.
\end{itemize}
Depending of relations between these characteristic times different spectra of CRs can be generated in the CND.

We notice that the value of diffusion coefficient $D$ in the CND
is, of course, unknown. The two limit  cases are possible. As it
was shown by \citet{kuls} magnetic fluctuations are damped in
a dense gas of molecular clouds because of ion-neutral friction,
that gives free CR escape from the clouds. On the other hand, a neutral gas in the
clouds is turbulized. This
turbulence excites forced magnetic fluctuations. 
As a result a spagetti like structure of magnetic field line is generated
inside the clouds and the energy of magnetic fluctuations is concentrated at a small correlation
length of the magnetic field, $L_{corr}$, which is much smaller than a size of the molecular cloud \citep[see][]{dog87,dog14,kiselev13}. These fluctuations prevent free escape of CRs from the clouds. As a result magnetized particles propagate along tangled magnetic field lines that can be described by diffusion with the coefficient
\begin{equation}
D\sim \beta\frac{c L_{corr}}{3}\,,
\label{d_coeff}
\end{equation}
 where $L_{corr}$ is the correlation length of turbulized magnetic field, and $\beta=v/c$. Below we estimate the value of
$D$ from solutions of kinetic equations and observational data.

\section{Processes of ionization and emission in the CND}
For the hadronic origin of the CND gamma rays,  their flux from the CND is
\begin{equation}
F_\gamma(E_\gamma,t) = \frac{n_Hc}{R^2_{GC}}\int\limits_0^Rr^2dr\int_E N(E,r,t)
\frac{d\sigma}{dE_\gamma}(E,E_\gamma)dE\,,
\label{fgamma}
\end{equation}
where $n_H$ is the  average hydrogen density in the gamma-ray
emitting region,  $V$ is its volume, $R_{GC} = 8$ kpc is the distance from Earth to the CND,  and $d\sigma/dE_\gamma(E,E_\gamma)$ is
the differential cross-section for gamma-ray production in
proton-proton collisions \citep{kamae}.

In the case of leptonic origin,  gamma rays can be generated by the
bremsstrahlung in the CND gas  and/or by the inverse-Compton
scattering of electrons on background photons,
\begin{eqnarray}
F_\gamma(E_\gamma) = \frac{n_Hc}{ R^2_{GC}}\int\limits_0^Rr^2dr\int N_e(E,r,t)
\left(\frac{d\sigma}{dE_\gamma}\right)_{br}dE + \nonumber \\
\frac{c}{ R^2_{GC}}\int\limits_0^Rr^2dr\int N_e(E,r,t) dE\int w(\epsilon,r)
\left(\frac{d\sigma}{dE_\gamma}\right)_{IC}d\epsilon\,,
\label{fgammae}
\end{eqnarray}
where $\left(\frac{d\sigma}{dE_\gamma}\right)_{IC}$ and
$\left(\frac{d\sigma}{dE_\gamma}\right)_{br}$ are  the
differential cross-section of inverse-Compton and bremsstrahlung
processes accordingly \citep[see][]{blu70}, $\epsilon$ is the energy of
soft photons (IR and optical) and $w(\epsilon)$ is their energy density.

The
ionization rate of hydrogen by the protons or electrons in this region can be
estimated from
\begin{equation}
\zeta =\int \sigma_HvN(E) dE \,,
\label{ionf}
\end{equation}
where $\sigma_H$ is the ionization cross-section of the molecular
hydrogen by  proton or electron impact whose equation is approximately \citep[see][]{spitzer}
\begin{equation}
\sigma_H=1.23\times
10^{-20}\frac{1}{\beta^2}\left[6.2+\lg\left(\frac{\beta^2}{1-\beta^2}\right)-0.43\beta^2\right]\mbox{cm$^2$}
\label{spitz}
\end{equation}
 for more accurate equations of cross-section \citep[see][]{rudd,tati03}.  

Protons produce also secondary electrons and positrons by $p-p$
collisions and by the knock-on process. The production rate of  secondary electrons is
estimated as
\begin{equation}
Q_e(E_e,{\bf r}, t) = n_H\int N(E,{\bf r}, t) v(E)\frac{d\sigma}{dE_e}dE\,,
\label{Qesec}
\end{equation}
where $\frac{d\sigma}{dE_e}$ is the differential cross-section for
electron and positron production. For the $p-p$ process the
cross-section is presented in e.g. \citet{kamae}.  The knock-on
cross-section has the approximated  form as \citep[see
e.g.][]{ginz}
\begin{equation}
 \frac{d\sigma}{dW}\simeq \frac{2\pi e^4}{m_ev^2W^2}
\end{equation}
where $v$ and $W$ are the velocity of a primary particle and the
energy of a secondary electron, respectively.  For more accurate
equations for the cross-section see \citet{dst75} and \citet{haya}.

The spectrum of secondary electrons in the CND is calculated from Eq.
(\ref{eq_state}) for $Q(E_e)$  in the form (\ref {Qesec}).

Relativistic electrons generate radio emission by the synchrotron
losses. The corresponding equation for synchrotron emission is
\citep[see for details][]{ginz}
\begin{equation}
\Phi(\nu,t)=\frac{\sqrt{3}e^3H}{m_ec^2R^2_{GC}}\int\limits_0^Rr^2dr\int\limits_0^{E_{max}}N(E_e,r,t)dE_e\frac{\nu}{\nu_c}\int
\limits_{\nu/\nu_c}^\infty K_{5/3}(x)dx
\label{radio}
\end{equation}
where $N(E_e)$ is the density of electrons with the energy $E_e$,
$K_\alpha(x)$ is the McDonald function and $\nu_c=
3eH\gamma^2/4�m_ec$, $\gamma$ is the Lorenz-factor of an electron.

Subrelativistic electrons and protons produce bremsstrahlung
X-ray emission. The equation for the  bremsstrahlung radiation is
\begin{equation}
F_{br}(E_x,t) =\frac{n_H}{R^2_{GC}}\int\limits_0^Rr^2dr\int N(E,r,t)v(E) \left(\frac{d\sigma}{dE_x}\right)_{br}dE\,,
\label{br}
\end{equation}
where
\begin{equation}
\left(\frac{d\sigma}{dE_x}\right)_{br}=\frac{8}{3}\frac{e^2}{\hbar
c}\left(\frac{e^2}{m_ec^2}\right)^2\frac{m_ec^2}
{\hat{E}E_x}\ln\left[\frac{(\sqrt{\hat{E}}+\sqrt{\hat{E}-E_x})^2}{E_x}\right]
\label{sbr}
\end{equation}
Here $\hat{E}=E_e$ for electrons and   $\hat{E}=(m_e/m_p)E_p$ for protons, where $m_p$ is the mass of the proton.

\section{Steady State Hadronic Model}
We start from the simplest case
of steady state model for protons. If the mean path length of
protons  equals or is longer than the CND radius,
$\sqrt{D\tau_{pp}}\geq R$ then we can use equation for the total
number of protons $\bar{N}(E)=\int\limits_0^Rr^2dr N(E,r)$ in the
CND. In this case the equation for the particle number $\bar{N}$ reads
as
\begin{equation}
\frac{d}{d E}\left(\frac{dE}{dt}\bar{N}\right)=\bar{Q}(E)
\label{eq_state1}
\end{equation}
where the energy loss term is taken in the form (\ref{sloss}), and $\bar{Q}(E)$ in the form (\ref{Q}).
The solution of Eq. (\ref{eq_state1}) is
\begin{equation}
\bar{N}(E) = \frac{1}{dE/dt}\int\limits_{E}^\infty \bar{Q}(E)dE \,.
\label{f2}
\end{equation}

As in Eq.(\ref{fgamma})  the flux of gamma rays with the energy $E_\gamma$ from the CND is
\begin{equation}
F_\gamma(E_\gamma)  = \frac{M_Hc}{4\pi R^2_{GC} m_p}\int \frac{\bar{N}(E)}{V}
\frac{d\sigma}{dE_\gamma}(E,E_\gamma)dE\,,
\label{fgamma1}
\end{equation}
where   $V$ is the CND volume, $M_H$ is the total mass of
hydrogen there.

\begin{figure}[h]
\begin{center}
\includegraphics[width=0.95\textwidth]{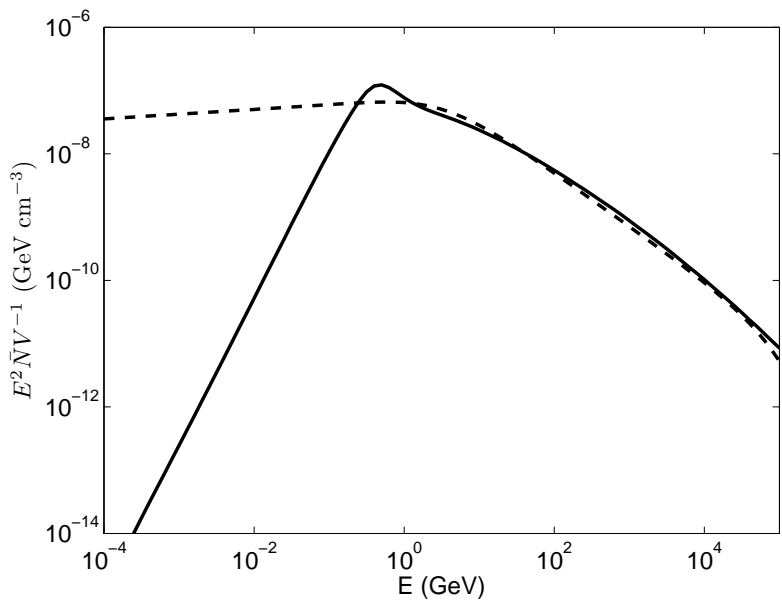}
\end{center}
\caption{The  energy spectrum of protons (solid line) derived from Eq. (\ref{f2}) and the data of \citet{masha11}
and \citet{katia12}. The injection spectrum is shown by the dashed line. }\label{fig:pspectra}
\end{figure}

We assume that the gamma-ray flux from the source 2FGL
J1745.6$-$2858 is generated  within the CND region. Then the
results depend only on the total  mass of the hydrogen within the
area covered by relativistic proton. The average spectrum of CR
protons $\bar{N}(E)/V$ derived for the  total mass of the CND and
the gamma-ray flux of J1745.6$-$2858 is shown in  Fig.
\ref{fig:pspectra}. From this figure one can see that the derived density of protons is below the estimates of \citet{masha11} in the relativistic energy range and strongly below the approximation of \citet{goto13} because of the ionization losses.

\begin{figure}[h]
\begin{center}
\includegraphics[width=0.8\textwidth]{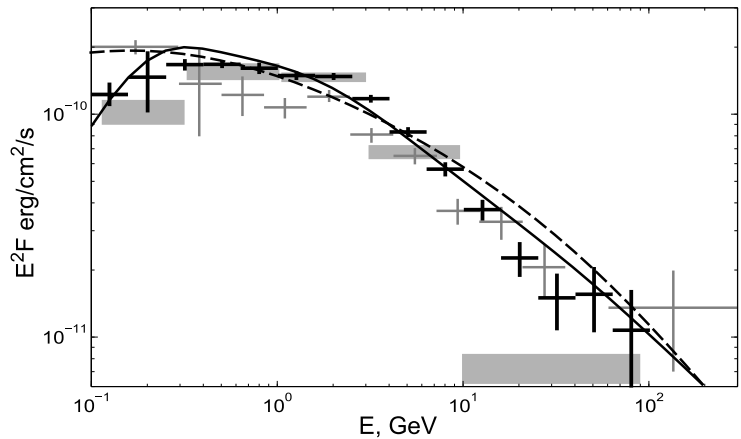}
\end{center}
\caption{Gamma-ray spectrum expected in the case of
hadronic origin of the gamma-ray emission (solid line) and in
the case of leptonic origin of the gamma-ray emission (dashed line).
Data points in form of gray crosses are from \citet{masha11}, data
points in form of gray bars are from the second Fermi catalog and data
points in form of black crosses are from this work.}
\label{fig:spectrum}
\end{figure}

The calculated gamma-ray spectrum produced by these protons is shown in Fig. \ref{fig:spectrum} by solid line.

For this spectrum of protons we can calculate from Eq.
(\ref{Qesec}) the production rate of secondary electrons.  In the steady state case the total number of secondary electrons can be estimated as
\begin{equation}
\bar{N}_e^{sec}(E_e) = \frac{1}{dE/dt}\int\limits_{E_e}^\infty \bar{Q}_e(E)dE \,,
\label{Nesec1}
\end{equation}
where the term $dE/dt$ describes the total energy losses of
electrons, which includes the ionization, bremsstrahlung,
synchrotron and inverse-Compton losses.

\begin{figure}[h]
\begin{center}
\includegraphics[width=0.8\textwidth]{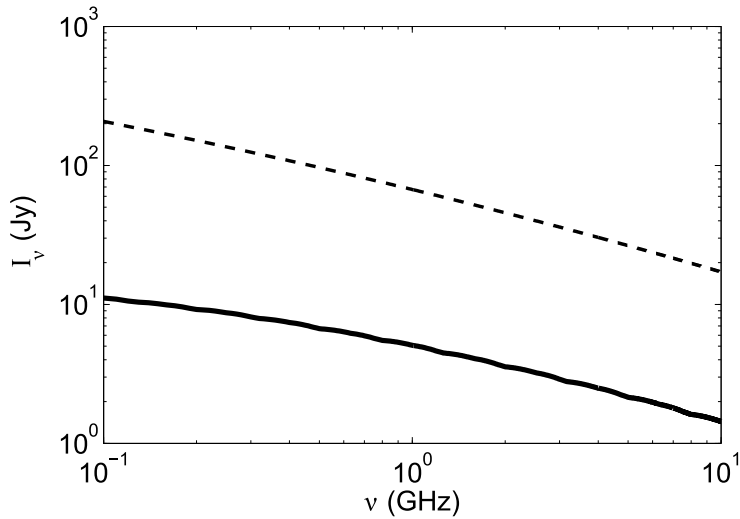}
\end{center}
\caption{Radio emission expected in the case of hadronic (solid
line) and leptonic (dashed line)
origin of the gamma-ray emission.} \label{fig:radio}
\end{figure}

The contribution of secondary electrons to the total gamma-ray
flux from the CND  is relatively low.
However, these electrons emit a flux of radio emission from the CND
region.  The radio spectrum in the CND is calculated from Eq. (\ref{radio}) and (\ref{Nesec1}) and it is shown in
Fig. \ref{fig:radio} by the heavy solid line.  The estimated radio
flux  at $\nu=1.5$ GHz is about 4.63 Jy.

The
ionization rate of hydrogen by the protons in the stationary model can be
approximated as
\begin{equation}
\zeta = \int \sigma_Hv\frac{\bar{N}(E)}{V} dE \,.
\label{ionf1}
\end{equation}
Our can calculations show that the ionization rate is about $4.6\times 10^{-15}$s$^{-1}$, of which $2.85\times 10^{-15}$s$^{-1}$ is produced by primary protons and $1.75\times 10^{-15}$s$^{-1}$ by secondary electrons.

From Eq.(\ref{br}) we calculated  X-rays flux in the range 20-40 keV generated by  bremsstrahlung of protons and electrons. The total flux is about $3.1\times 10^{-13}$erg cm$^{-2}$s$^{-1}$ ,  protons generate $2.27\times 10^{-13}$erg cm$^{-2}$s$^{-1}$ and secondary electrons - $0.83\times 10^{-13}$erg cm$^{-2}$s$^{-1}$.   The spectral index of bremsstrahlung radiation in this range  is $-1.53$.

\section{Transient Injection of Protons}

Here we investigate another limit case of proton injection, namely the case of
transient injection. As  in \citet{masha11} we
 assume that
the injection of protons happened $t$ years ago, and  the
gamma-ray flux from the CND at present is just as observed from 2FGL
J1745.6$-$2858.

For the transient injection the source function $Q(E,r,t)$ is taken in the form
\begin{equation}
Q(E,r,t)=A(E)\delta({\bf r})\delta(t)
\end{equation}
where $A(E)$ is given by Eq. (\ref{q_pwlaw}) that gives  $A(E)=A_0E^{-\gamma}$ in the relativistic  energy range, and
$A(E)=A_0'E^{-(\gamma+1)/2}$ in the subrelativistic energy range.

Using the Green function of Eq.(\ref{eq_state}) as derived by \citet{syrov} the spectrum can be presented as
\begin{equation}
N(E,r,t)=\frac{A_0E^{-\gamma}}{(4\pi Dt)^{3/2}}e^{-t/\tau_{pp}}\exp\left(-\frac{r^2}{4Dt}\right)
\label{sp_rel}
\end{equation}
 in the relativistic energy range and
\begin{equation}
 N(E,r,t)=\frac{A_0'\sqrt{E}}{(4\pi Dt)^{3/2}}\exp\left(-\frac{r^2}{4Dt}\right)\left(at+E^{3/2}\right)^{-(\gamma+1)/3}
\label{sp_subrel}
\end{equation}
in the subrelativistic energy range. Here according to Eq.(\ref{iot})
the rate of ionization energy losses in the subrelativistic range is presented as $dE/dt=-a/\sqrt{E}$.

 We
analyse the  two cases when  
\begin{equation}
\sqrt{4D\tau_{pp}}<R~~~~ \mbox{and}~~~~ \sqrt{4D\tau_{pp}}>R\,. 
\label{ineq}
\end{equation}
\begin{enumerate}
\item In the first case the protons do not leave the CND region and are absorbed there for the time $\tau_{pp}$. The diffusion coefficient in the dense CND gas  is determined by a spagetti-like structure of magnetic field lines and has the form (\ref{d_coeff}), the case of \citet{dog87}. As it follows from the first inequality (\ref{ineq}), the correlation length of the magnetic field $L_{corr}<10^{18}$ pc.  The proton spectrum has a maximum in the subrelativistic energy range at (see  Eq. (\ref{sp_subrel}))  
\begin{equation}
E_m\simeq (at)^{2/3}\,.
\end{equation} 
where $t$ is a current time, $t<\tau_{pp}$. The evolution of proton spectrum for the diffusion coefficient $D=3\times 10^{27}$ cm$^2$s$^{-1}$ is shown in Fig. \ref{fig05},
\begin{figure}[h]
\begin{center}
\includegraphics[width=0.8\textwidth]{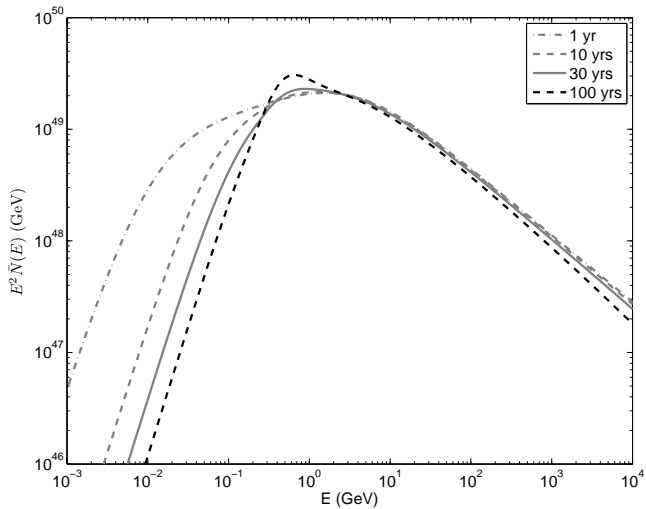}
\end{center}
\caption{Evolution of the spectrum of protons injected by a point-like source. The diffusion coefficients in this case is $D=3\times 10^{27}$ cm$^2$s$^{-1}$.} \label{fig05}
\end{figure}
\item In the second case a significant fraction of protons leave the CND, and only a small fraction of them interact with the CND gas. For the CND conditions this case corresponds to the inequality, $L_{corr}\geq R$. It means that CRs propagate through the CND almost without "scattering"   \citep[the case of][]{kuls}. The residence time of particles in the CND is  about $R/\beta c$, which equals $\sim 10$ yr for relativistic particles. The maximum position in the proton spectrum shifts to low energies as
\begin{equation}
E_m\sim \left(a\frac{R}{\beta c}\right)^{2/3}\,. \label{em}
\end{equation}
The density of particles inside the CND is determined by  particle propagation  (diffusion)  in the medium surrounding the CND.
\end{enumerate}

As it follows from numerical calculations of  \citet{kiselev13} the relation between the correlation length of magnetic fluctuations $L_{corr}$ and a size of  molecular cloud depends on many parameters of the cloud, such as the strength of large scale magnetic field, the frequency of collision between ionized and neutral particles etc.

\begin{figure}[h]
\begin{center}
\includegraphics[width=0.8\textwidth]{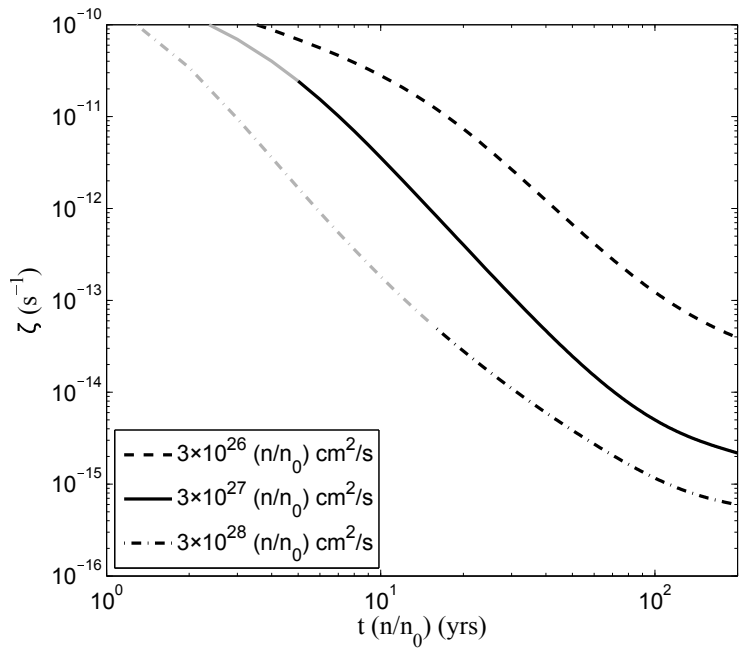}
\end{center}
\caption{The ionization rate $\zeta$ at the center of CND as a
function of  $t$ and $D$. } \label{ioni}
\end{figure}

From Eqs. (\ref{ionf}) and (\ref{spitz}) we calculated the
ionization rate $\zeta$ in  CND center for  different values of $D$. The ionization rate $\zeta$  at the center of CND as a
function of $t$ and $D$ is shown in Fig. \ref{ioni}. The gray parts of lines show parameter regions where the diffusion approximation does not work, i.e. when the lifetime of particles is shorter than the characteristic time of "scattering" by magnetic fluctuations.
One can see that the calculated ionization rate is close to derived by \citet{goto13} for
the CND, $\zeta\simeq 1.2\times 10^{-15}$ s$^{-1}$, if $D\geq 3\times 10^{27}$ cm$^2$ s$^{-1}$. In this case a part of  primary protons escape from the CND. 

\begin{figure}[h]
\begin{center}
\includegraphics[width=0.95\textwidth]{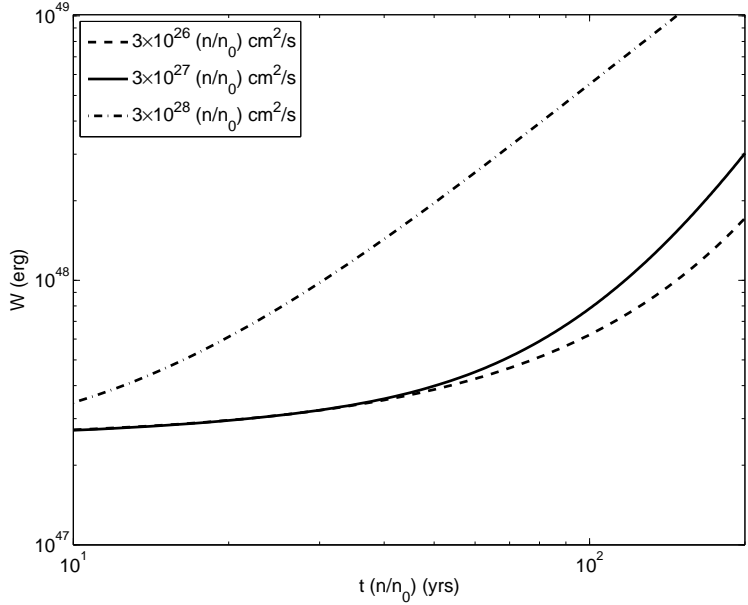}
\end{center}
\caption{The energy release needed for the observed intensity of gamma-ray emission
from the CND.}
\label{q}
\end{figure}

The total gamma-rays flux from the CND region consists of the three components, one of which is produced by $p-p$ collisions and two others by bremsstrahlung and inverse Compron losses of secondary electrons. Then from Eqs. (\ref{eq_state}), (\ref{fgamma}), (\ref{fgammae}), (\ref{Qesec}) and (\ref{sp_rel}) we can estimate the energy release in the form of primary relativistic protons needed for the observed gamma-ray flux from the CND.
Its value is shown in Fig. \ref{q} for different time $t$ and different diffusion coefficients $D$.

\begin{figure}[h]
\begin{center}
\includegraphics[width=0.8\textwidth]{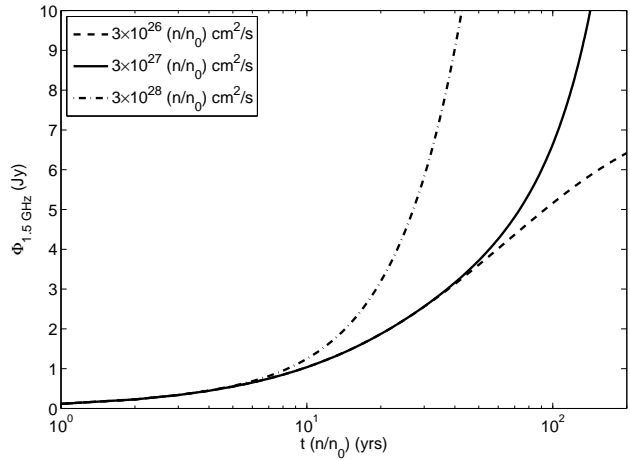}
\end{center}
\caption{Synchrotron radio emission from the CND region at the frequency $\nu=1.5$ GHz for different time momenta $t$ and different $D$. } \label{rad_em}
\end{figure}

The flux of synchrotron emission generated by secondary electrons can be calculated from Eq. (\ref{radio}). The flux of synchrotron radio emission from the CND region at the frequency $\nu=1.5$ GHz for different time  $t$ and different $D$ is shown in Fig. \ref{rad_em}.
An interesting characteristic  of the CND radio emission is an increase of the flux with time. The reason is that the lifetime of electrons emitting at the frequency 1.5 GHz is quite long, about $10^4$ yr. Therefore, for large enough $t$ and $D$ (that requires a high enough injection rate of primary protons in the past, see Fig. \ref{q}) a significant fraction of electrons, generated at  initial stages,  escape from the CND and form a halo around this region. Just this "old" population of electrons emitted in the past is responsible for a high radio flux observed at the current time $t$.

\begin{figure}[h]
\begin{center}
\includegraphics[width=0.8\textwidth]{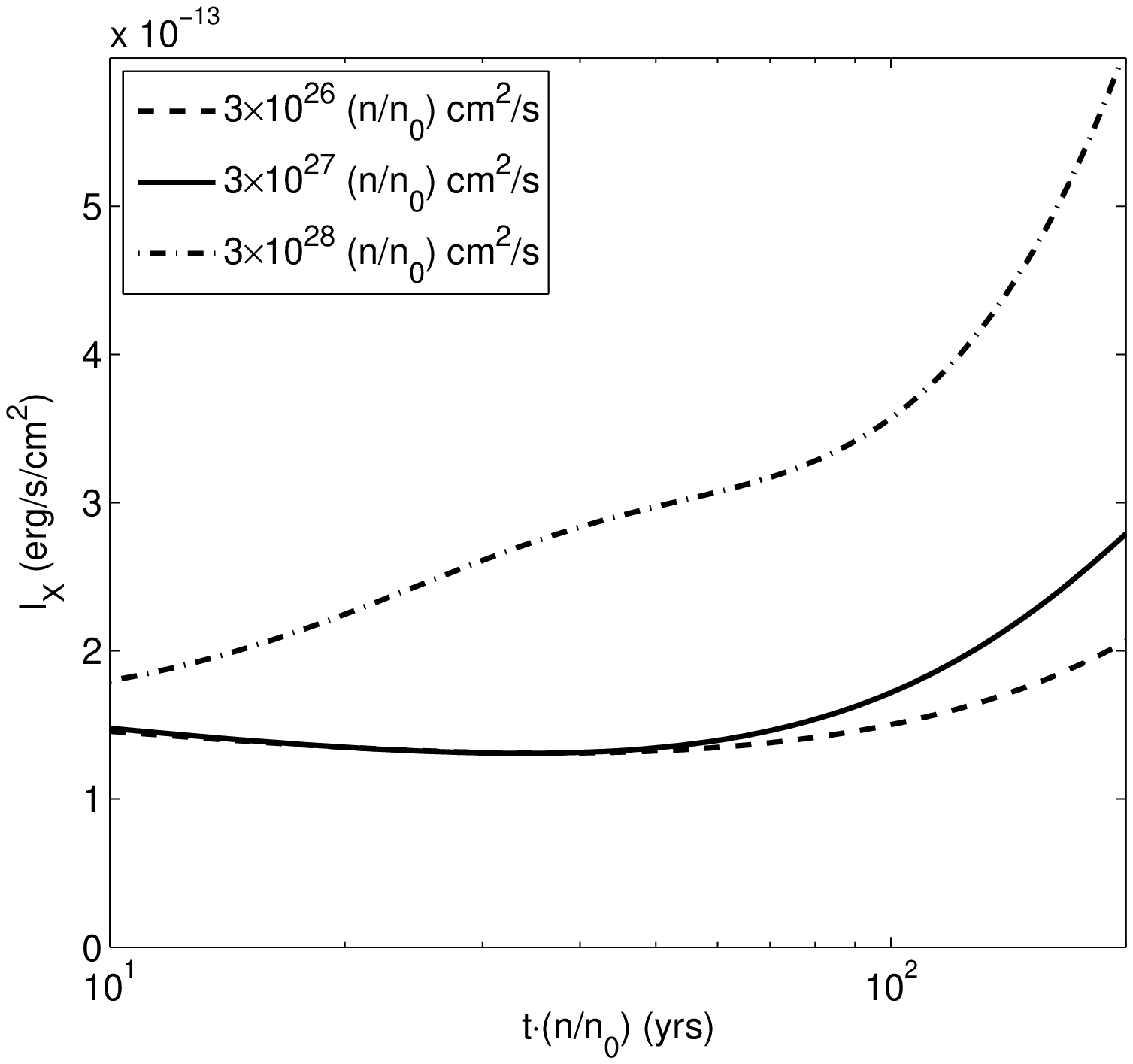}
\end{center}
\caption{X-ray emission from the CND region in the range $20 - 40$ keV for different time momenta $t$ and different $D$. } \label{rad_x}
\end{figure}

We expect that secondary electrons generated by proton collisions with the gas provide also a flux of X-rays from the CND. The expected flux of hard X-ray photons in the range $20-40$ keV is shown in  Fig. \ref{rad_x}.  As one can see the  flux increases  also  for large $t$. To interpret this effect we notice that the CND X-ray emission is generated  by the three processes: by inverse bremsstrahlung of primary protons, and by bremsstrahlung  and inverse Compton  of secondary electrons. If the diffusion coefficient $D$  is relatively small, this emission from the CND is mainly produced by inverse bremsstrahlung of protons. However, for large enough $D$ a contribution from inverse Compton of $\sim 100$ MeV electrons which escape from the CND and fill a region around the CND, becomes significant. In this case the area of X-ray emission is more extended than the CND region. 

\begin{figure}[h]
\begin{center}
\includegraphics[width=1.0\textwidth]{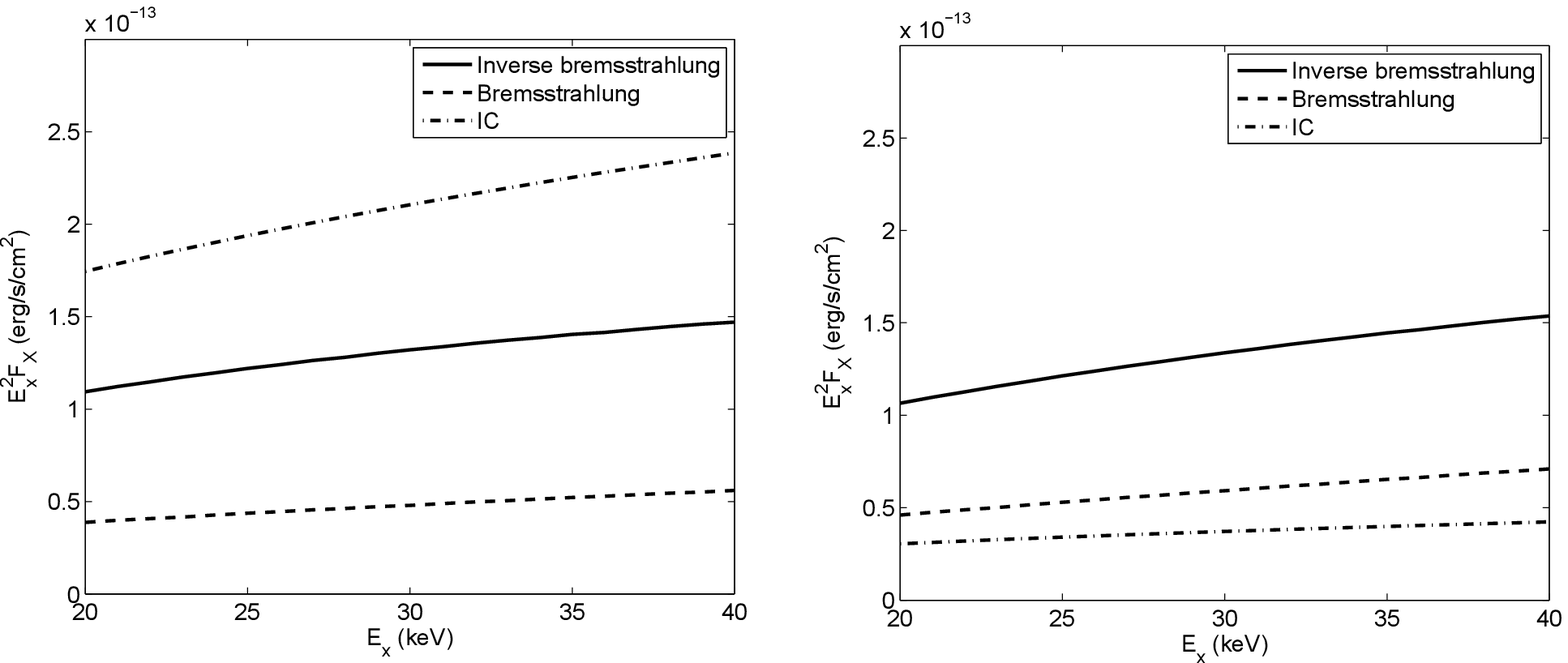}
\end{center}
\caption{Components of X-ray emission at the time $t=100$ yr for the diffusion coefficient:  $D=3\times 10^{28}$cm$^2$s$^{-1}$  (left panel), and $D=3\times 10^{27}$cm$^2$s$^{-1}$  (right panel) } \label{comp}
\end{figure}

As an example we presented spectra  of X-ray components for the diffusion coefficients: $D=3\times 10^{27}$cm$^2$s$^{-1}$  (Fig. \ref{comp}, right panel) and $D=3\times 10^{28}$cm$^2$s$^{-1}$  (Fig. \ref{comp}, left panel), the current time is $t=100$ yr in both cases.

As it follows from our analysis the model of transient proton injection is able to reproduce characteristics of nonthermal emission from the CND and the rate of hydrogen ionization there if the proton eruption occured about hundred years ago (see e.g. Fig. \ref{ioni}). We notice that a high past activity of Sgr A*  follows from the temporal variations of the 6.4 keV
iron line  observed in the direction of GC
molecular clouds \citep[see][and references therein]{inu09,ponti10,terrier,nob11,clavel13} who found that about 100 years ago the energy release of Sgr A* was several orders of magnitude higher than at present. May it be that the gamma-ray flux from the CND,  hydrogen
ionization there and the time variable flux of the 6.4 keV line at present
 have a common origin, namely, a past activity of Sgr A*?

\section{Hydrogen Ionization by  CR Electrons}
In the case of leptonic origin of gamma rays from the CND the
expected  ionization rate of the neutral hydrogen can be estimated
in the same way as done in the previous section. The spectrum of
electrons $N_e(E)$ can be estimated from Eq. (\ref{eq_state}) for the production spectrum in the form
(\ref{Q}) and (\ref{q_pwlaw}).

The energy loss term, $dE/dt$,  includes the
ionization, bremsstrahlung,  synchrotron and inverse-Compton
losses \citep[see corresponding equations in][]{haya, ginz}. We
note that for the average density $n = 4\times 10^5$ cm$^{-3}$,
average magnetic field $H = 1$ mG and density of soft optical and
IR photons $w_{op} = w_{IR} = 10^4$ eV/cm$^3$, the bremsstrahlung
losses dominate up to energies about of 25 GeV. For the ionization
losses  we used the ESTAR database \citep{star} which provided the
data up to 10 keV electron energy.  Below 10 keV we used the approach for ionization
losses from \citet{dalg}.

In the case of leptonic origin  gamma rays can be generated by
bremsstrahlung in the interstellar gas  and/or by  inverse-Compton
scattering of electrons on background photons, as described by Eq. (\ref{fgammae}).

For observed gamma-ray flux of the source  2FGL J1745.6$-$2858 we
 derived from Eq. (\ref{fgammae}) the electron spectrum, and
estimated the hydrogen ionization rate provided by these electrons
in the CND.

The  gamma-ray spectrum derived from the leptonic model is shown
in Fig. \ref{fig:spectrum} by dashed line. For the accepted parameters gamma rays are
mostly produced by bremsstrahlung. The ionization
rate provided by the electrons is $6.3\times 10^{-14}$ s$^{-1}$. One can see that the  ionization rate expected in the leptonic mode is
higher than derived by \citet{goto13} from the H$_3^+$ data, $1.2\times 10^{-15}$ s$^{-1}$.
This circumstance makes the leptonic interpretation problematic,
unless the electrons providing the gamma rays and the
hydrogen ionization have different origin. 

The estimated X-ray flux from the CND generated by the electron bremsstrahlung is about $5.5\times 10^{-12}$ erg cm$^{-2}$s$^{-1}$ in the range $20-40$ keV. The spectral index of this emission is -1.37.

The expected radio emission produced by these electrons from the
CND region is shown in  Fig. \ref{fig:radio} by the dashed and
dash-dotted lines. The value of radio flux at 1.5 GHz is 62 Jy. This value is more than one order of magnitude larger than the estimate of 5 Jy obtained by \citet{yusef14} that also makes leptonic interpretation problematic.

\section{Discussion}
Our analysis shows that the hydrogen ionization rate derived in
the framework of the  stationary hadronic model of gamma rays in the CND is
about the same as the value observed by \citet{goto13,goto14}. If these
gamma rays  are produced by relativistic electrons, then the
expected ionization rate is higher than observed.  Besides, as
follows from the recent analysis of the 1.5 GHz flux  from the
inner 2 arcmin region within Sgr A* \citep{yusef14} its value is about 5 Jy, just
as expected in the hadronic model. In this respect, the hadronic
model looks more attractive than the leptonic model which gives at 1.5 GHz the radio flux from the CND  $\geq 50$ Jy.
Further analysis of the radio emission from
the GC and determination of radio counterparts of the gamma-ray
source may restrict the set of the models and reveal the nature of
the source.  We notice that radio data are decisive for the model
choice because both models predict an exact number of relativistic
electrons.

Parameters of ionization and emission derived in the framework of nonstionary (transient) hadronic model are time variable and depend on characteristics of Sgr A* past activity. As it follows from our analysis the ionization rate of hydrogen $\simeq 1.2\times 10^{-15}$s$^{-1}$ and radio flux $\sim 5$ Jy as obtained by \citet{goto13} and \citet{yusef14} for the CND region can be obtained in this model if the diffusion coefficient there is about $3\times 10^{27}$ cm$^2$s$^{-1}$ and the proton eruption occured about hundred years ago (see Figs. \ref{ioni} and \ref{rad_em}).  An interesting circumstance of the nonstationary model is a significant contribution of electrons generated at the initial stages of activity into the fluxes of  radio emission and  hard X-rays.  The expected flux of X-rays in the energy range $20 - 40$ keV is about $1.5\times 10^{-13}$ erg cm$^{-2}$s$^{-1}$ with the spectral index about -1.5. These electrons having relatively long lifetime  escape from the CND region and generate in the surrounding medium radio flux by synchrotron and a flux of  X-rays by inverse Compton. In this case the emission region is more extended than the CND. 

 It is interesting to notice that a transient
release of energy by Sgr A* 100 years ago was derived from the observed
temporal variations of the 6.4 keV and hard X-ray continuum fluxes
in the GC region \citep[see][and references
therein]{inu09,ponti10,terrier,nob11}. It was concluded that just
about 100 years ago Sgr A* was a source  of hard X-rays with the
luminosity $\sim 10^{39}$ erg s$^{-1}$. We can hypothesize that, if the transient hadronic model describes properly processes of emission and ionization in the CND, then there may be a
connection between a past proton eruption and a past X-ray activity of Sgr A*.

At the end we would like to mention that new generation gamma-ray telescopes may
provide more information about  the  source 2FGL J1745.6$-$2858 and
resolve it with higher accuracy than the \fermi. For example, if  the telescope
GAMMA-400 is able to detect the gamma-ray flux of this source, then with the claimed angular resolution about $0.01^\circ$ \citep{gamma400a, gamma400b}
it is able to resolve the source within 1.5 pc.

\section*{Acknowledegments}
The authors thank Farhad Yusef-Zadeh for his very useful comments
and the  unpublished radio data which we used in the paper. The authors thank Nikolay Topchiev and Sergei Suchkov for description of GAMMA-400 properties.
 DOC is supported in parts by the RFFI
grant 12-02-31648, the LPI Educational-Scientific Complex and
Dynasty Foundation. DOC and VAD acknowledge support from the RFFI
grants  15-52-52004, 15-02-02358, 15-02-08143. KSC is supported by the  GRF Grants of the
Government of the Hong Kong SAR under HKU 7010/13P. DOC, KSC, and
VAD acknowledge support from the International Space Science
Institute to the International Team "New Approach to Active
Processes in Central Regions of Galaxies".

\end{document}